\documentclass[a4paper,10pt]{article}
\usepackage[utf8]{inputenc}

\usepackage{amssymb}
\usepackage{graphicx}
\usepackage{bm}
\usepackage{amsmath}
\usepackage{longtable}
\usepackage{lscape} 
\usepackage{color}
\usepackage[labelfont=bf,labelsep=period,justification=raggedright]{caption}

\title{Escaping the Tragedy of the Commons through Targeted Punishment}
\author{Samuel Johnson\\
\small{Warwick Mathematics Institute, and Centre for Complexity Science,}\\
\small{University of Warwick, Coventry CV4 7AL, United Kingdom.}\\
\small{E-mail:  S.Johnson.2@warwick.ac.uk}}

\date{}

\begin{document}

\maketitle

%%%% Abstract text to be placed here %%%%%%%%%%%%
\begin{abstract}
Failures of cooperation cause many of society's gravest problems. 
It is well known that cooperation among many players faced with a social dilemma can be maintained
thanks to the possibility of punishment, but achieving the initial state of widespread cooperation is often much more difficult. 
We show here that there exist strategies of `targeted punishment' whereby a small number of punishers
can shift a population of defectors into a state of global cooperation.
The heterogeneity of players, often regarded as an obstacle, can in fact boost the mechanism's effectivity.
We conclude by outlining how the international community could use a strategy of this kind to combat climate change.
\end{abstract}
%%%%%%%%%%%%%%%%%%%%%%%%%%%

%%%%%%%%%% Insert the texts which can accomdate on firstpage in the tag "fmtext" %%%%%

%\begin{fmtext}
\section{Introduction}
When Svante Arrhenius enunciated his greenhouse law in 1896, atmospheric concentration of CO$_2$
stood at its highest in over half a million years
-- about $300$ ppm \cite{Arrhenius,Etheridge}. 
It has now surpassed $400$ ppm \cite{NatGeo}.
%According to the Intergovernmental Panel on Climate Change, 
Our continued failure to avoid the well-known consequences 
of global warming is not rooted in some technical impossibility, but in a lack of international cooperation \cite{IPCC,Burning}.
It is a classic example of the `tragedy of the commons', as popularised by Garrett Hardin through the metaphor of 
herdsmen with access to common pasture land: each can always prosper individually by adding another head of cattle to his 
herd, but eventually this leads to overgrazing and ruin for all \cite{Hardin,Lloyd}.
Other instances include overfishing, deforestation, and many kinds of pollution.
The solution advocated by Hardin was ``mutual coercion, mutually agreed upon'',
which is usually taken to mean 
coercion by a Hobbesian central authority \cite{Radkau}.
Some argue that an alternative option is to privatise the commons \cite{Smith}, although
the coercion is still implicit in the assumption
%\linebreak
%\end{fmtext}
%%%%%%%%%%%%%%% End of first page %%%%%%%%%%%%%%%%%%%%%
%\maketitle
%\noindent
that property rights can be enforced \cite{Choi}.
Ironically perhaps, it is in local communities with access to some resource similar to Hardin's common pasture land 
where self-organization to cooperate has often been documented \cite{Ostrom_book,Lansing}. 
And indeed, such cases usually involve rules, mutually agreed upon, and enforced by the possibility of some form of 
punishment \cite{Ostrom_Science}.

In game theory, the tragedy of the commons is seen as a Nash equilibrium, where no rational agent cooperates despite its 
being the strategy which would maximise collective payoff if adopted by all players \cite{Nash}. 
Since cooperation is, nevertheless, pervasive in nature and society,
%Since there are, nevertheless, many instances of cooperation in nature and society, 
much theoretical and empirical work has gone into understanding why this might be 
so \cite{Axelrod_book,Maynard_Smith,Nowak_Evolving,Reciprocity_rev,Camerer,Nowak_rules,Capraro}.
The focus has been on behaviour in the face of `social dilemmas' -- situations where there is some communal benefit in choosing a 
cooperative strategy, but also a temptation to defect (not cooperate).
In one-to-one games such as the {\it prisoner's dilemma}, a strategy of conditional cooperation can be individually 
advantageous if the game is iterated and players are able to remember each other \cite{Axelrod_book,Grujic}. 
A better model for commons management, however, is the {\it public goods game} \cite{Kollock,Moreno}. Each
player can choose how many tokens to put into a common pot which multiplies the total amount by some factor 
(greater than one and smaller than the number of players),
and redistributes the result equally among all players.
Many experiments with humans have shown that cooperation (i.e. adding to the pool) can be enhanced by allowing players to
punish defectors,
%(those who fail to cooperate), 
despite the punisher incurring a cost for 
doing so \cite{Reciprocity_rev,Nowak_Winners}.

One aspect of 
social 
%cooperative
dilemmas which is not usually taken into account in theoretical studies is the heterogeneity
of players \cite{Nowak_Evolving}: even in lab experiments where the small number of subjects are all students, 
significantly different attitudes to cooperation 
are found \cite{Reciprocity_rev}. 
In cases where the players are nation states, the differences are much larger. 
When it comes to tackling global warming, for instance, 
the heterogeneity in gross and per capita emissions, vulnerability to climate change, 
dependence on fossil fuels, historical responsibility, technical and financial ability to adapt, and many other relevant variables is
widely regarded as confounding the problem \cite{IPCC,Burning}.

When public goods experiments are run in the lab, with the same group of subjects playing iteratively, 
cooperation tends to be high at first and 
gradually dwindle thereafter, possibly as cooperators become frustrated with the behaviour of defectors \cite{Reciprocity_rev}. 
Allowing 
players to punish defectors from the start in such settings can discourage would-be defectors and maintain cooperation. 
In the real world, however, 
the problem is often not just one of maintaining cooperation, but of achieving it in an environment of almost ubiquitous defection.
For instance, in a society with very little corruption, maintaining this happy state is relatively 
easy, since anyone attempting to break the rules would swiftly be identified and punished.
In an environment of entrenched corruption, however, there are usually too many defectors and too few resources to change
the state of affairs \cite{Corruption}.

It is often assumed that punishment -- and indeed positive incentives -- must be seen as fair, and there is some 
evidence that unfair or inconsistent punishment fails to maintain cooperation in lab experiments
\cite{vanProoijen_Inconsistent}. But in situations of widespread defection, the punishing capacity of would-be punishers (usually a 
subset of cooperators), if applied equally to all defectors, can be too dilute to have any effect.
%I 
Here we use a simple model to show how, in such situations, there exist strategies of `targeted punishment' which 
punishers can adopt in order to escape the tragedy of the commons and bring about universal cooperation. Far from being an obstacle,
the existence of heterogeneity among the players 
contributes to the strategy's effectivity, and may, perhaps, serve to 
%can be harnessed to afford objectivity to the strategy, and perhaps 
assuage any feelings of unfairness.

\section{Results}

\subsection{Maintaining vs achieving cooperation}

% New description of model:
Let us consider a set of $N$ players faced with a social dilemma of some kind. At any given moment, each player can choose either 
to cooperate or to defect. Depending on the details of the situation, a player will perceive a net payoff associated with 
each strategy. Let us call the difference of these perceived payoffs $H_i$ for player $i$. Thus, if $i$ has all the relevant 
information, and is entirely selfish and rational, she will cooperate if $H_i>0$ and defect is $H_i<0$.
We shall consider, however, that the degree to which these assumptions hold can be captured by a `rationality' parameter $\beta$,
in such a way that $i$ has, at each time step $t$, a probability $P_i$ of cooperating and a probability $1-P_i$ of defecting, where
\begin{equation}
P_i=\frac{1}{2}[\tanh(\beta H_i)+1].
\label{eq_Pi}
\end{equation}
This sigmoidal 
form coincides with the transition probabilities for the spins in an Ising model, and for the neurons in 
a Hopfield neural network \cite{Marro_book}. Behaviour is completely random if $\beta=0$, and becomes deterministic (perfectly rational)
when $\beta\rightarrow\infty$.
The need to take this feature into account is suggested by work
%A well-known limitation of classical game theory is the assumption of perfect rationality, and several approaches
in evolutionary game theory and behavioural economics which has highlighted the importance of somewhat stochastic or bounded rationality
\cite{May_noise,Nowak_noise,Bounded_rationality}.

What form shall we choose for $H_i$? We are interested in situations where,
in the absence of interaction with the rest of the population,
most of the players are predisposed to defect.
However, these predispositions can be heterogeneously distributed.
%However, the predispositions are not necessarily the same for all players.
%-- in fact, it is precisely the heterogeneity in these predispositions which is often regarded as a major obstacle in international 
%tragedies of the commons.
%, but which, as we shall see, can also provide paths to global cooperation.
Let us assume, with no loss of generality,
that the sequence $i=1,2,... N$ positions the players in order of their predisposition, from most to least intrinsically cooperative.
For simplicity, let us consider that the predisposition $h_i$ of player $i$ is given by 
the linear expression
$h_i=-(i-2)/(N-2)$. Thus, for any $N$, 
the first player is the only one with a slight tendency to cooperate,
%($h_1=1/(N-2)$), 
the second one has no inherent tendency,
%($h_2=0$),
and each successive player has a greater tendency to defect than the previous one, down to the last with $h_N=-1$.
In addition to this individual effect, each player can be influenced by the others. For instance, let us assume that a certain number of 
players $n_p$ 
have each a capacity $\pi$ to punish defecting players they consider at fault, of which there are $n_f$. The total punishment befalling 
a defector among the $n_f$ is then $p_i=\pi n_p/n_f$. The balance of payoffs for a given player considered at fault is now
\begin{equation}
H_i=p_i+h_i=\pi\frac{n_p}{n_f}-\frac{i-2}{N-2}.
%H_i=h_i+p_i=-\frac{i-2}{N-2}+\pi\frac{n_p}{n_f}.
\label{eq_Hi}
\end{equation}
(Note that $n_p$ and $n_f$ can change with time, although for clarity we refrain from making this explicit.)
This is also the balance $H_i$ for players who are cooperating but who would become at fault if they were to defect
(with the small adjustment that, since such a player would presumably not punish herself,
she has $p_i=\tilde{n}_p/\tilde{n}_f$, where 
$\tilde{n}_p$ and $\tilde{n}_f$ are the values of $n_p$ and $n_f$ that there would be if this player defected).
Meanwhile, for players not considered at fault irrespectively of their strategies, $p_i=0$.

This simple model captures the features of social dilemmas required to illustrate how targeted punishment can work,
without sacrificing generality by going into the details of a given game.
However, the parameter $\pi$ could be adjusted to describe,
for instance, the public goods game with specific punishment costs.
%A well-known limitation of classical game theory is the assumption of perfect rationality, and several approaches in 
%evolutionary game theory and behavioural economics have highlighted the importance of somewhat stochastic or bounded rationality
%\cite{May_noise,Nowak_noise,Bounded_rationality}. We shall see that rationality, as captured by the parameter $\beta$,
%also plays an important role here.

Which players can punish, and whom should they punish? In many real situations, it is only possible for cooperators
to punish defectors. In this case, $n_p$ is equal to the number of cooperators, $n_c$, at any given time.
For now we shall focus on this case, although the possibility of defectors also punishing other defectors is discussed 
below.\footnote{There is no reason, in principle, why defectors should not be able to retaliate by punishing the punishers.
However, since the focus here is on situations where all players recognise the collective benefits of cooperation, as in the case of 
global warming, 
%I will
we shall leave this possibility unexplored.}
As to who should be punished, this is in fact the only ``rule'' that the community has freedom to determine -- or, more precisely, 
that those in a position to punish can determine. The simplest (and arguably fairest) 
rule would be for all defectors to be punished. In this case, $n_f$ 
is equal to the number of defectors, $n_f=N-n_c$.

\begin{figure}[t!]
\begin{center}
\includegraphics[scale=0.55]{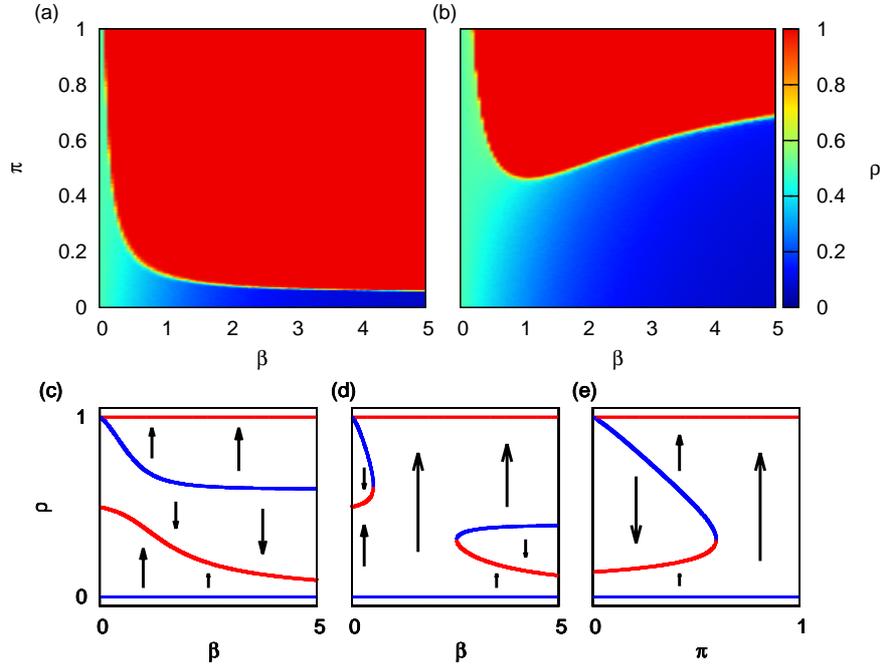}%{fig_xscat.eps}%{fig_1_sketch.eps}
\end{center}
\caption{
(a) Stationary proportion of cooperators, $\rho$, for a range of rationality, $\beta$, and punishment, $\pi$, from Monte Carlo simulations 
of the model when all cooperators punish all defectors, and initially all  $N=200$ players cooperate. 
(b) As before, but now all players initially defect.
(c) Fixed points of the dynamics against $\beta$, when $\pi=0.4$; stable fixed points are depicted in red, unstable ones in blue.
(d) As in (b), but with $\pi=0.6$.
(e) Fixed points of the dynamics against $\pi$, when $\beta=2.5$. (The fixed-point analysis is described in Methods.)
%Phase diagrams.
}
\label{fig_1}
\end{figure}

We run computer simulations of the situation described above for $N=200$ players (roughly the number of countries in the world)
and compute the average proportion of cooperators, $\rho=n_c/N$, once a stationary state has been reached. Figure \ref{fig_1}
shows this proportion on a colour scale for a range of the two parameters, $\beta$ (rationality), and $\pi$ (punishment).
In panel (a), we see that for almost all parameter combinations global cooperation is obtained. However, there is an 
important detail: for these simulations, we have set the initial strategy of every player to `cooperate'. The lesson we can learn, 
therefore, is that in these conditions global cooperation can be maintained once it has been achieved. But what about if the 
initial strategies are all set to `defect'? In Figure \ref{fig_1}b we show the results for this case.
There is now a much smaller region of global cooperation, requiring significantly higher levels of punishment $\pi$ than are necessary simply 
to maintain cooperation. Interestingly, while a certain degree of rationality $\beta$ is needed to achieve cooperation, thereafter 
the minimum punishment enabling cooperation increases with rationality, implying that some degree of randomness in the selection of 
strategies is globally beneficial.

What is happening here? To gain a better understanding of the phenomenon, we consider the 
{\it fixed points}
%(also, somewhat misleadingly, sometimes called {\it equilibrium points}) 
of the dynamics. These are values $\rho^*$
such that, when $\rho=\rho^*$, the subsequent value to which the system naturally evolves is, again, $\rho^*$. A fixed point 
can be either stable or unstable: if a small deviation from $\rho^*$ would tend to return the system to $\rho^*$, it is stable;
whereas it is unstable if random fluctuations around this point are amplified and the system driven to some other value of $\rho$.
In Methods we analyse the fixed points and their 
stability, and show the results for three different combinations of parameters in the bottom panels of Figure \ref{fig_1}.
Figure \ref{fig_1}c corresponds to a level of punishment $\pi=0.4$. The lines show the fixed points as functions of $\beta$, 
with stable fixed points plotted in red and unstable ones in blue. Arrows show the direction in which the system will tend to evolve
depending on the value of $\rho$ (away from unstable fixed points and towards stable ones). 
First of all, we observe that global defection ($\rho=0$) is always unstable, while global cooperation ($\rho=1$) is stable for any 
$\beta>0$. There are two further fixed points, one stable and one unstable. If the system begins with sufficient cooperators that 
$\rho$ is above the unstable one, it will evolve towards global cooperation. However, if the initial $\rho$ is below this, 
evolution will be towards the other stable fixed point.
%Thus, if $\beta$ is small behaviour is quite random, and nearly $50\%$ of players 
%will cooperate. With increasing rationality, however, this number decreases towards zero.
This explains the difference between 
the top two panels, where global cooperation is observed when all players begin cooperating, but not when they start off defecting.
Figure \ref{fig_1}d shows a situation of higher punishment, $\pi=0.6$. There are now still regions of $\beta$ for which the 
stable fixed point at low $\rho$ acts as a trap when all players begin defecting; but an interval has appeared in which there 
is an uninterrupted path from $\rho=0$ to $\rho=1$. This corresponds to the region in the top right panel where cooperation can be
observed at this $\pi$ for intermediate values of $\beta$. Finally, in Figure \ref{fig_1}e we set $\beta=2.5$ and plot the fixed 
points against $\pi$. Again we see that, for $\pi$ below a certain value, there is a stable fixed point at low $\rho$ which acts as a 
trap, while high enough $\pi$ will ensure global cooperation irrespectively of initial conditions.

%The {\it coexistence of phases} we observe in this model, such that different global behaviour can be reached from the same 
%parameter values, is one of the hallmarks of a discontinuous (first order) phase transition -- in contrast to the continuous 
%(second order) transition which arises in the standard Ising model \cite{Ising_scholarpedia}.

%This phenomenology -- a feature of any system in which a given opinion, behaviour or strategy can become 
%entrenched by virtue of its very universality -- will motivate our search for non-trivial rules leading from widespread defection to 
%global cooperation.

\subsection{Paths to cooperation}

\begin{figure}[t!]
\begin{center}
\includegraphics[scale=0.35]{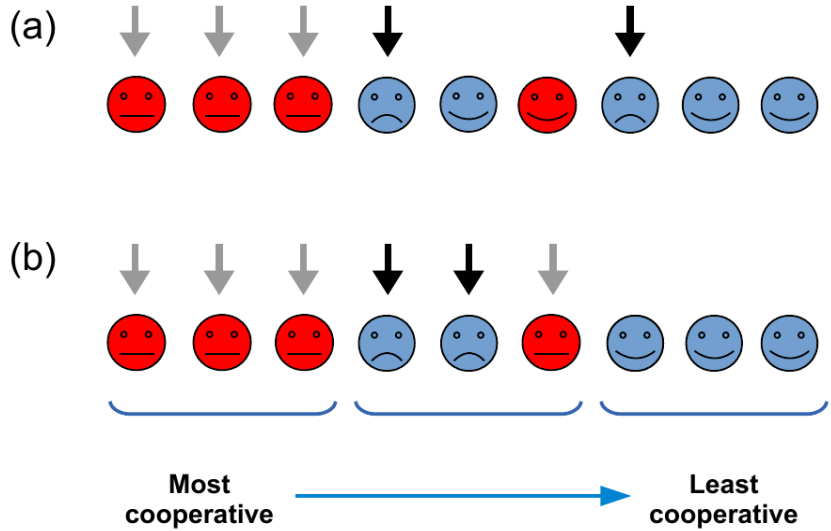}%{fig_a.eps}%{fig_xscat.eps}%{fig_1_sketch.eps}
\end{center}
\caption{
Diagrams illustrating the two strategies of targeted punishment described in the main text: (a) is the `single file' strategy, 
and (b) is the `groups' strategy with groups of size $\nu=3$ and a threshold $\theta=2/3$. Players are arranged from most to least 
inherently cooperative; those currently cooperating are shown in red and those defecting in blue. A black arrow indicates a defector
who is considered at fault (and therefore liable to be punished) according to the strategy, 
while a grey arrow signals a cooperator who would be at fault if she were defecting.
}
\label{fig_a}
\end{figure}

\begin{figure}[t!]
\begin{center}
\includegraphics[scale=0.55]{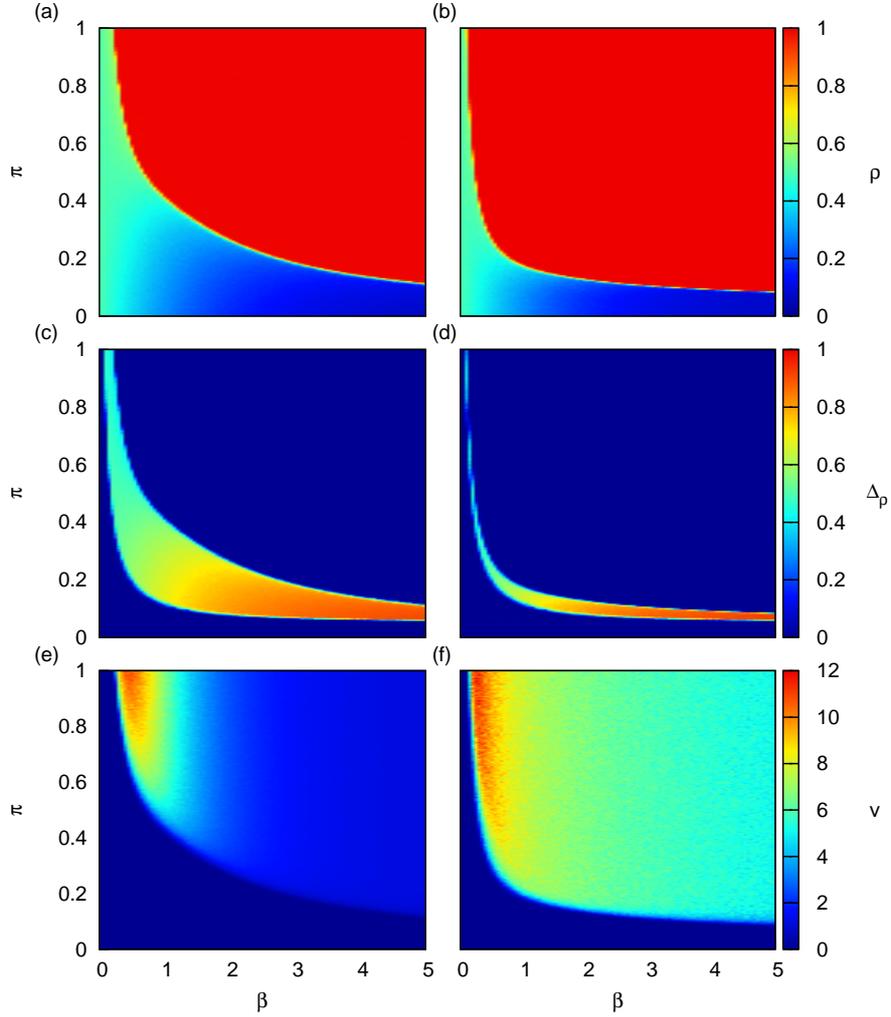}%{fig_2.eps}%{fig_xscat.eps}%{fig_1_sketch.eps}
\end{center}
\caption{
(a) As Figure \ref{fig_1}b (all players initially defect), but now the `single file' strategy is applied.
%the rule is that a defector is only deemed 
%at fault if the one before her in the ordering is cooperating.
(b) As in Figure \ref{fig_1}b, but under the `groups' strategy with $\nu=10$ and $\theta=80\%$.
%players are assigned to $20$ groups of size $\nu=10$, and a defector is deemed at fault if a at least
%a proportion $\theta=80\%$ of the previous group is cooperating.
(See the main text and Figure \ref{fig_a} for descriptions of these strategies.)
(c) Difference
%, $\Delta_\rho$, 
between Figure \ref{fig_1}a (all players initially cooperate) and Figure \ref{fig_2}a.
(d) Difference
%, $\Delta_\rho$, 
between Figure \ref{fig_1}a and Figure \ref{fig_2}b.
(e) Speed $v=N/\tau$, where $\tau$ is the number of time steps required to achieve global cooperation, for the situation in Figure \ref{fig_2}a.
(f) Speed $v$ for the case of Figure \ref{fig_2}b.
}
\label{fig_2}
\end{figure}

As noted above, players with the ability to punish others 
have the freedom to decide whom to punish. It may seem fairest to punish 
all defectors equally, but when these are numerous this approach dissipates the total punishing capacity.
Consider, instead, the following rule. A defecting player $i$ is only deemed at fault at time $t$ if the one immediately before her
in the ordering, player $i-1$, cooperates at time $t$. 
This rule, which we shall refer to as the `single file strategy', is illustrated in Figure \ref{fig_a}a.
According to this view, the number of players considered at fault, $n_f$, will 
be smaller than the total number of defectors when these are in the majority, while the scenario becomes identical to the 
previous one when almost all the players cooperate.
In Figure \ref{fig_2}a we show the results for simulations in which punishers adopt this strategy.
As in Figure \ref{fig_1}b, all players initially defect.
The region of global cooperation is now significantly larger than in Figure \ref{fig_1}b: harmony can 
be achieved at much lower values of punishment $\pi$, particularly if rationality $\beta$ is high.
Because at any one time only a very small number of defectors are deemed at fault, even a low level of punishment 
is sufficient to make them cooperate. As each new player switches strategy, it passes on the 
burden of responsibility to another one down the line,
resulting in a cascade of defectors becoming cooperators.
%The process is therefore a cascade thanks to the conditional commitments made at the outset.
A secondary effect is that, as the ranks of cooperators grow, the total 
punishment they are able to inflict increases, although as we show in Supplementary Material this is not essential for the 
mechanism to work.

The single file strategy
allows for 
global cooperation to ensue from widespread defection in situations where this would not have been possible with equal allocation of 
punishment. Thus, selecting only certain players for culpability can provide an escape route from the
tragedy of the commons. However, this is not necessarily the best rule to ensure such an outcome; in fact, there are regions at low 
$\pi$ and $\beta$ where, according to Figure \ref{fig_1}, global cooperation is sustainable, yet not achievable via this route.
So consider now the following arrangement,
which we can call the `groups strategy'.
Players are allocated to groups of size $\nu$, such that players $i=1,...\nu$ belong 
to the first group, $i=\nu+1,...2\nu$ to the second, and so forth. A defector 
belonging to group $m$
is deemed to be at fault at time $t$ if and only if 
at least a proportion 
$\theta$ of the players in group $m-1$ 
cooperate at time $t$.
This strategy is illustrated in Figure \ref{fig_a}b.
In Figure \ref{fig_2}b we show 
simulation results for this scenario, with $20$ groups of $\nu=10$ players and a threshold of $\theta=80\%$, where, as before, all 
players begin defecting. (To set the process off, players in the first group are always considered at fault if they defect.)
An even greater region of parameter space now corresponds to global cooperation, leaving only very low levels of $\pi$ and $\beta$ out 
of reach. For a given number of players, $N$, and set of inherent tendencies, $h_i$, there will be optimal rules,
or `targeted punishment strategies',
which come closest to ensuring global cooperation for any values of rationality and punishment.
(Note that the single file strategy is an instance of the more general groups strategy when $\nu=1$ and $\theta=100\%$.)

Figures \ref{fig_2}c and \ref{fig_2}d show the difference, $\Delta_\rho$, between the maximum density of cooperators 
achievable (i.e. when all players initially cooperate), and the results of Figures \ref{fig_2}a and \ref{fig_2}b, respectively.
About $12\%$ of the parameter space shown corresponds to situations where cooperation is possible but not 
achievale via the single file strategy. For the groups strategy, however, little over $3\%$ of the potential parameter space remains
out of reach.
Another aspect to take into account when comparing punishment strategies is the speed with which 
cooperation can be achieved. Figures \ref{fig_2}e and \ref{fig_2}f show the quantity $v=N/\tau$ for each rule, where $\tau$ 
is the number of time steps required to achieve cooperation. In most of the parameter range, 
cooperation is achieved sooner with $\nu=10$ than with $\nu=1$.
%the strategy based on groups provides a faster route to cooperation than the single file strategy.

As remarked above, in many real situations it is only the cooperators who are seen as having the ability or legitimacy 
to punish defectors. However, defectors too could, in principle, punish other defectors, even if this may be regarded as 
somewhat unfair.
For instance, in many societies criminals pay value added tax on their purchases, thereby contributing
indirectly to the penal system.
%For instance, in society criminals use money to make purchases and pay value added tax, thereby contributing
%indirectly to the penal system. 
In Supplementary Material we perform the same analysis for the case in which $n_p=N$; that is, all players contribute to 
the punishment of defectors. The dynamics is qualitatively similar to the situation in which only cooperators can punish, the main 
difference being that global cooperation can, unsurprisingly, ensue from lower levels of punishment per player.
If, on the other hand, only a fraction $a$ of cooperators were to punish, the situation would be as in Figures \ref{fig_1} and \ref{fig_2}
after rescaling $\pi\rightarrow a\pi$.

The situations thus far examined involve a predisposition to cooperate, $h_i$, with a specific functional form;
and the punishing strategies assume that their precise ordering is known.
In Supplementary Material we relax these constraints by adding a Gaussian noise to $h_i$, and randomly switching the ordering 
of $25\%$ of players with randomly chosen counterparts. We find that both punishing strategies described above are quite robust to 
these changes: the single file strategy is the most robust at high levels of both punishment and rationality, while the 
%opposite is true 
groups strategy is superior
at low values of these parameters.

%%% New discussion

\section{Discussion}

Punishment has been shown to maintain cooperation in many social dilemma settings \cite{Reciprocity_rev,Nowak_Winners},
and it is generally assumed that such punishment should be fair \cite{vanProoijen_Inconsistent}.
However, in situations of entrenched defection, the society's punishing capacity can become too dilute to have any effect if applied equally 
to all defectors.
The message of this paper is that even in such situations there can exist strategies of `targeted punishment' which allow  
a few initial punishers to shift a large number of defectors into a state of global cooperation. 
%The heterogeneity in players' natural predispositions to cooperate, which is often regarded as exacerbating the problem,
%can contribute to the efficacy of targeted punishment.
%The paths to cooperation described above 
%Such paths to cooperation would seem, however,

The paths to cooperation described above would seem to rely heavily on the possibility of punishment.
Since punishing a defector presumably has some cost, such an act in itself constitutes a kind of cooperation.
This is not necessarily a problem, given that humans and governments alike are wont to 
engage in
%``altruistic punishment''
``costly punishment''
in a variety of settings \cite{Reciprocity_rev,Nowak_Winners,Sanctions}.
%Yet 
But, in any case, punishment
%this 
is only one 
potential mechanism which might give rise to a term $p_i$ with the characteristics we have here assumed. For instance, a determining 
factor in human behaviour often seems to be the anticipation of how one's choices might affect those of others \cite{Capraro}. 
We know that our 
recycling, voting or travelling by bicycle will have little impact on the world per se, but we may rationally engage in these 
activities in the hope that others will follow suite. If the rules of the game have been set up in such a way that our actions 
determine whether the next player in line will be expected to honour her conditional commitments, what seemed like a grain of 
sand in the desert becomes a grain of sand in an avalanche. If one imagines all eyes turned towards the single player whose turn 
it is to cooperate -- or to the single small group of such players -- it is easy to see how one might be more inclined to cooperate
than in a world of distributed responsibility.

One could argue that establishing the initial ordering would be an obstacle of similar magnitude to achieving cooperation 
directly. This may be the case when the players are alike in all respects. But an acknowledgement of heterogeneity 
might break this symmetry in a way acceptable to all, especially if there exist objective measures to establish, say, the 
effort each player would have to make to cooperate. Furthermore, a player fairly well inclined to cooperate but deterred by the 
mass of less well-predisposed companions might happily adopt an early position in the hope that the mechanism may bear fruit;
while staunch defectors can leave the burden of responsibility to others by being placed further down the line, with the 
knowledge that they would only be called upon to participate if global cooperation were nigh.
In any case, if the tool for 
convincing players to cooperate is some form of punishment, only the punishers need agree on whom to punish at any given time.

It is worth reflecting that much social organization as we know it is in fact achieved though an implicit arrangement of targeted punishment.
Even the most despotic tyrants cannot personally punish all dissenters. But if they can exert power over a small group of underlings, who in 
turn manage their subordinates, and so on down a hierarchical pyramid, top-down control can occur. Similarly, most of us are subject to the 
judging gazes of only our immediate friends and neighbours, yet this can be enough to ensure conformity to various social conventions.

\subsection{Targeted punishment in practice}

Could a strategy of targeted punishment be implemented by the international community to escape from global tragedies of the commons, such 
as anthropogenic climate change? Ideally, countries might sign up voluntarily to small groups, each of which would in turn be allocated 
a position in an ordering.
Alternatively, the ordering and groupings could follow automatically from some objective measure, such as income per capita.
Small nations already making significant yet unsung 
progress, or particularly vulnerable ones, 
could use their early positioning to draw 
attention to their situations in the hope of having a wider effect. Others 
may initially welcome the temporary lifting of responsibility by signing up to a group further down the line, but find themselves obliged 
to cooperate once the spotlight came their way. Finally, even the biggest polluters would run out of excuses once a majority of other groups
were cooperating.
Some combination of sanctions and incentives could be arranged, although the mere fact of the whole world's eyes being focused
on a small number of defectors at any one time might prove a sufficient inducement in many cases. 
We have already tried signing up to commitments, enshrining these in law, privatising the commons through carbon trading schemes --
yet yearly global CO$_2$ emissions are now about $30\%$ higher than when the Kyoto Protocol was adopted \cite{IPCC,Burning}. 
Perhaps it is time for a 
%more strategic 
new approach.

\section{Methods}

According to Eqs. (\ref{eq_Pi}) and (\ref{eq_Hi}), the probability that player $i$ will cooperate at time step $t+1$ is
\begin{equation}
P_i(t+1)=\frac{1}{2}\tanh \left[\beta \left(\pi\frac{n_p(t)}{n_f(t)} -\frac{i-2}{N-2} \right)\right] +\frac{1}{2},
\label{eq_Pi_t1}
\nonumber
\end{equation}
where $n_p(t)$ and $n_f(t)$ are the numbers of punishing and punishable players, respectively, at time $t$.
If $\rho_t=n_c(t)/N$ is the proportion of cooperating players at time $t$,
let us define the expected proportion of cooperating players at time $t+1$: $G(\rho_{t})=\overline{\rho_{t+1}}$ (this is an expected value
in the sense that the average of $\rho_{t+1}$ over many independent realizations of the system will converge to $\overline{\rho_{t+1}}$).
We can then write
$$
G(\rho_{t})=\langle P_i(t+1) \rangle,
$$
where $\langle \cdot \rangle$ stands for an average over all players.
For the case where $n_p=n_c$ and $n_p=N-n_c$ (all defectors are punished by, and only by, all cooperators), this becomes
\begin{equation}
G(\rho_{t})=\frac{1}{2}\left\langle  \tanh \left[\beta \left(\pi\frac{\rho_t}{1-\rho_t} -\frac{i-2}{N-2} \right)\right]\right\rangle
+\frac{1}{2}.
\label{eq_G}
\end{equation}
Any value $\rho^*$ such that $G(\rho^*)=\rho^*$ will be a fixed point of the dynamics.
Fluctuations around $\rho^*$ will tend to dampen out if
\begin{equation}
\frac{d G(\rho_{t})}{d \rho_{t}}\vline_{\rho^*}\in (-1,1),
\label{eq_dG}
\end{equation}
whereas if the absolute value of the derivative is larger than one, the fixed point will be unstable, since even an infinitesimal 
fluctuation will drive the system to a different state. The bottom panels of Figure \ref{fig_1} are obtained by solving 
Eqs. (\ref{eq_G}) and (\ref{eq_dG}) numerically.
More generally, for any $\rho_t$, it is possible to determine whether the system can be expected to evolve towards more or fewer 
cooperators by the sign of $G(\rho_t)-\rho_t$.

\section{Acknowledgment}

Many thanks to Nick Jones, Iain Johnston, Janis Klaise, Lucas Lacasa and Fiamma Mazzocchi Alemanni for useful conversations and suggestions.

%%%%%%%%%% Insert bibliography here %%%%%%%%%%%%%%

\end{document}

\maketitle

%\section*{Supplementary Materials}

\setcounter{section}{0}

In the main text we describe two strategies of targeted punishment which punishers can adopt when most of the players are defecting:
\begin{itemize}
 \item Single file: A defector is considered at fault if and only if the player immediately before her in the ordering is currently cooperating.
 \item Groups: A defector is considered at fault if and only if at least a fraction $\theta$ of the players making up the 
 group immediately before hers in the ordering are currently cooperating.
\end{itemize}
The ordering should arrange players in descending order of their net perceived payoff $h_i$ -- 
that is, in increasing order of their temptation to defect.
We show that by focusing their punishment on defectors considered at fault according to the rule adopted, punishers can 
shift a population of defectors towards global cooperation in many situations where attempting to punish all defectors 
would have no appreciable effect. But the generality of these results is limited by two assumptions: that the number of punishers is 
proportional to the number of cooperators; and that the perceived payoffs of players follow a linear form, which is known to 
punishers. In this appendix we relax both assumptions and find that the effectivity of targeted punishment does not
depend strongly on such considerations.

%\section*{Constant punishment}
\section*{Constant punishment}

In the main text, we consider only scenarios in which the number of punishers is equal to the number 
of cooperators. If it were, in fact, proportional to the number of cooperators, this would simply involve a rescaling of the punishment
parameter, $\pi$. But what if all players were punishers, irrespectively of their individual state of cooperation? 
Figure S\ref{fig_ap1} shows the situations corresponding to Figure 1
%\ref{fig_1}
of the main text, with the difference that now 
the number of punishers is $n_p=N$ at all times. As in the case where only cooperators punish, there is a large region of parameter space
where cooperation is sustainable, but not achievable when punishment is diluted among all defectors.

\setcounter{figure}{0}
\renewcommand{\figurename}{Figure S}

\begin{figure}[ht!]
\begin{center}
\includegraphics[scale=0.55]{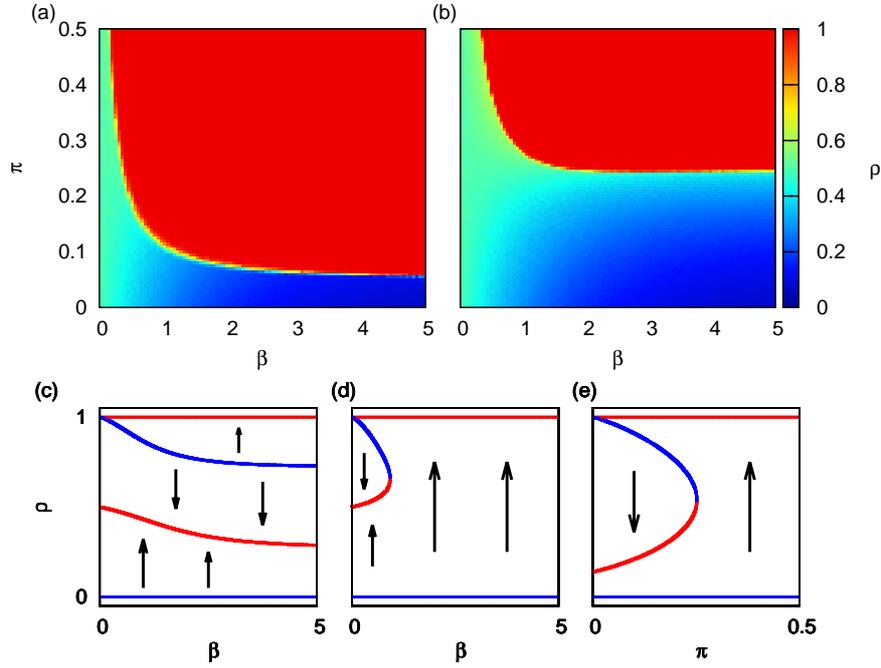}%{fig_xscat.eps}%{fig_1_sketch.eps}
\end{center}
\caption{
As Figure 1 of the main text, but with the number of punishers being equal to the number of players: $n_p=N$.
(a) All players initially cooperate. (b) All players initially defect.
Bottom panels: Stability diagrams when $\pi=0.2$ (c), $\pi=0.3$ (d), and $\beta=2.5$ (e).
%Like Figure 1 of main text, but with all punish rule.
%Left: As the top right panel of Figure \ref{fig_1} (all actors initially defect), but now the rule is that a defector is only deemed 
%at fault if the one before her in the ordering is cooperating.
%Right: As the left panel, but now actors are assigned to $20$ groups of size $\nu=10$, and a defector is deemed at fault if a at least
%a proportion $\theta=80\%$ of the previous group is cooperating. (See the main text for a description of these strategies.)
}
\label{fig_ap1}
\end{figure}

\begin{figure}[ht!]
\begin{center}
\includegraphics[scale=0.55]{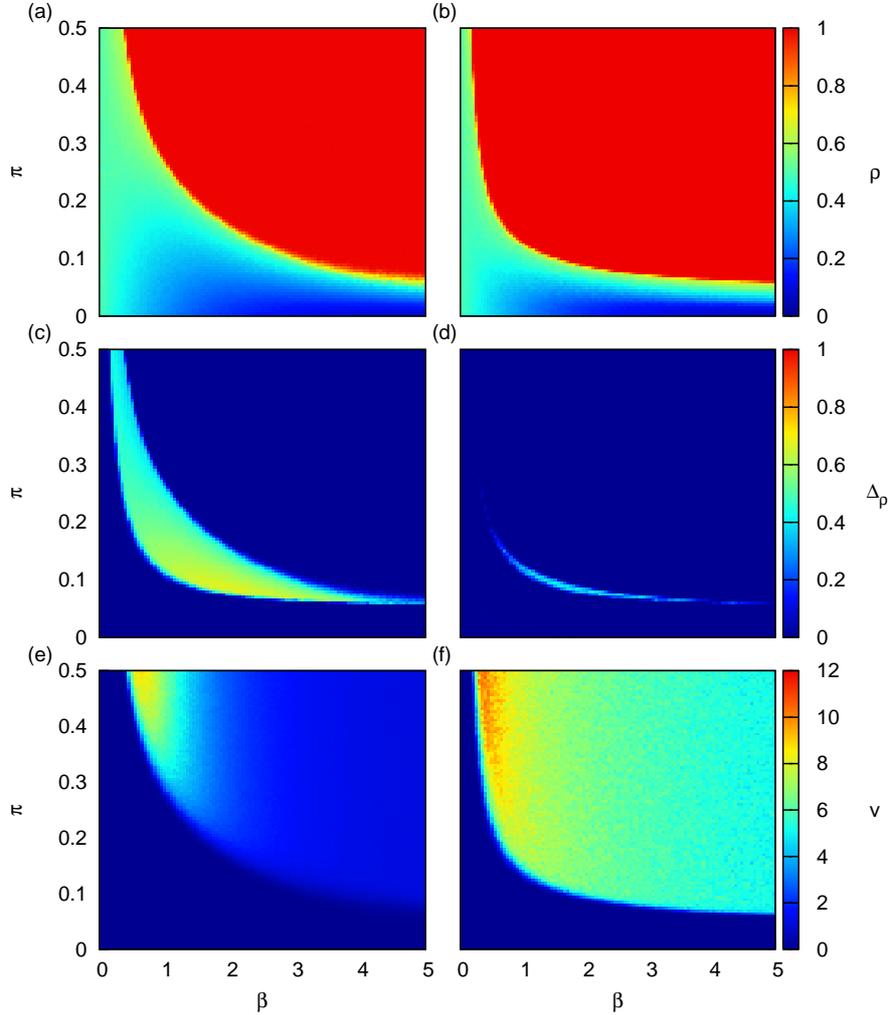}%{fig_ap1_tp12_SI.eps}%{fig_ap1_SI.eps}%{fig_xscat.eps}%{fig_1_sketch.eps}
\end{center}
\caption{
As Figure 3
%\ref{fig_2}
of the main text, but with the number of punishers being equal to the number of players: $n_p=N$.
(a) As Figure S\ref{fig_ap1}b (all players initially defect), but now the `single file' strategy is applied.
%the rule is that a defector is only deemed 
%at fault if the one before her in the ordering is cooperating.
(b) As in Figure S\ref{fig_ap1}b, but under the `groups' strategy with $\nu=10$ and $\theta=80\%$.
%players are assigned to $20$ groups of size $\nu=10$, and a defector is deemed at fault if a at least
%a proportion $\theta=80\%$ of the previous group is cooperating.
%(See the main text and Figure \ref{fig_a} for descriptions of these strategies.)
(c) Difference
%, $\Delta_\rho$, 
between Figure S\ref{fig_ap1}a (all players initially cooperate) and Figure S\ref{fig_ap1_tp}a.
(d) Difference
%, $\Delta_\rho$, 
between Figure S\ref{fig_ap1}a and Figure S\ref{fig_ap1_tp}b.
(e) Speed $v=N/\tau$, where $\tau$ is the number of time steps required to achieve global cooperation, 
for the situation in Figure S\ref{fig_ap1_tp}a.
(f) Speed $v$ for the case of Figure S\ref{fig_ap1_tp}b.
}
\label{fig_ap1_tp}
\end{figure}

The targeted punishment strategies described above are able to bring about global cooperation 
in large regions of the parameter space, as can be seen in Figure S\ref{fig_ap1_tp} -- this figure corresponds to Figure 3
%\ref{fig_2}
of the main 
text, with the difference that here $n_p=N$, instead of $n_p=n_c$. Note that, if we wished to consider a situation where a fraction 
$a$ of players where punishers, it would suffice to rescale the punishment parameter as $\pi\rightarrow a\pi$ in Figures S\ref{fig_ap1}
and S\ref{fig_ap1_tp}.

%\section*{Robustness to noise}
\section*{Robustness to noise}

\begin{figure}[ht!]
\begin{center}
\includegraphics[scale=0.55]{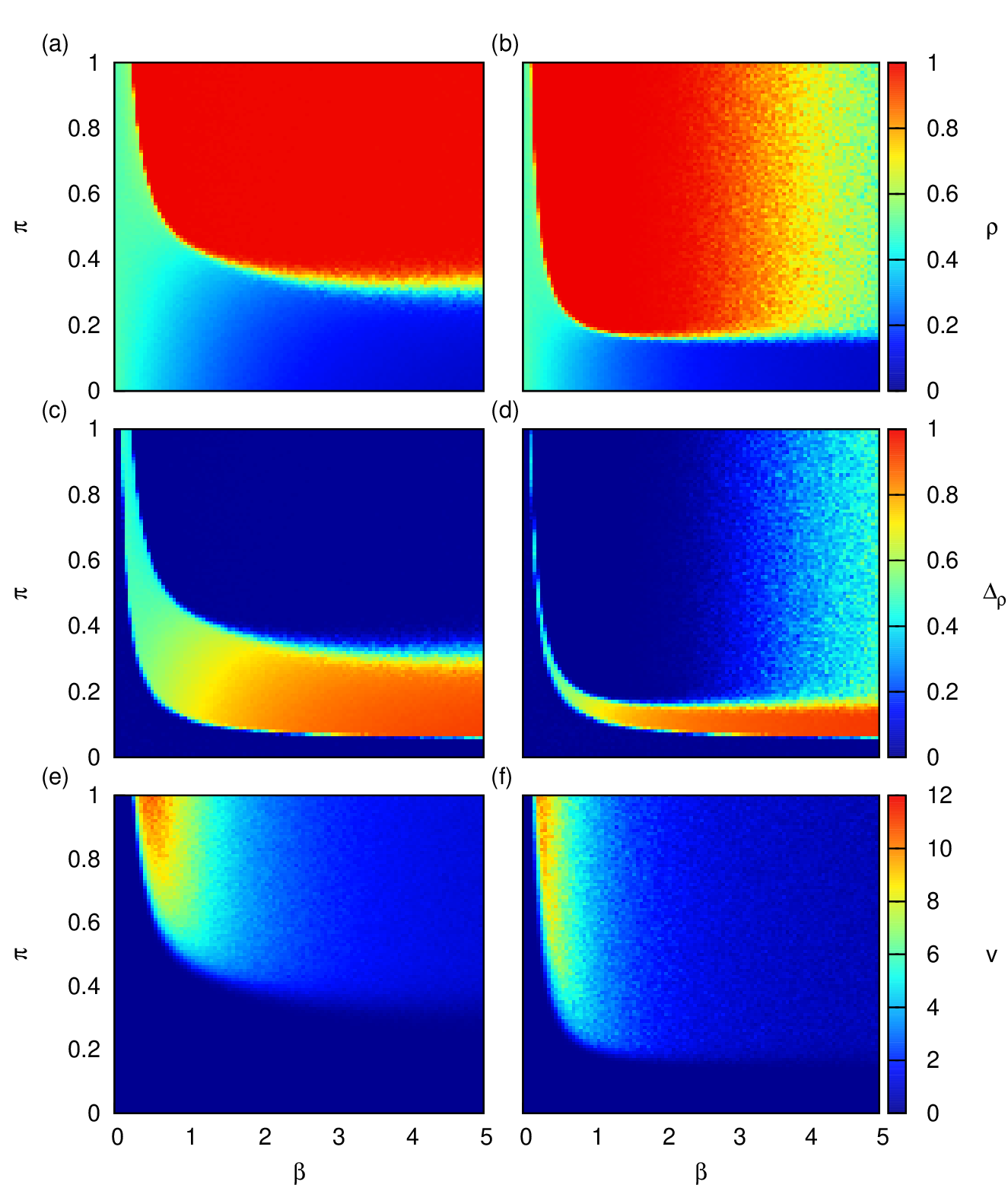}%{fig_noise_SI.eps}%{fig_2_SI.eps}%{fig_xscat.eps}%{fig_1_sketch.eps}
\end{center}
\caption{
As Figure 3
%\ref{fig_2}
of the main text, after the payoffs have been corrupted with a noise drawn from a Gaussian of mean zero and variance $\sigma^2=1$;
and a random proportion $f=25\%$ of players have had their positions in the ordering switched with other random players.
Punishment strategies are `single file' in the panels on the left [(a),(c) and (e)], and `groups' in those on the right [(b),(d) and (f)].
(a) and (b) Stationary proportion of cooperators, $\rho$.
(c) and (d) Difference between maximum $\rho$ maintainable [as displayed in Figure 1(a) of the main tex], and the results of
Figures S\ref{fig_noise}a and S\ref{fig_noise}b, repectively.
(e) and (f) Speed $v=N/\tau$, where $\tau$ is the number of time steps required to achieve global cooperation, 
for the situations in Figures S\ref{fig_noise}a and S\ref{fig_noise}b, respectively.
}
\label{fig_noise}
\end{figure}

In the main text we consider situations where player $i$'s payoff, in the absence of punishment, is $h_i=-(i-2)/(N-2)$,
and players are arranged in descending order of $h$. In real situations, punishers may not know 
the payoff perceived by player $i$. We therefore corrupt this setting with two sources of noise to gauge the strategies' 
robustness to this imperfect knowledge. We now consider that $i$'s payoff is $h_i=-(i-2+\eta_i)/(N-2)$, where the variables 
$\eta_i$ are drawn from a Gaussian with mean zero and variance $\sigma^2$. We also reshuffle the ordering, by choosing a 
fraction $f$ of the $N$ players randomly, and switching their positions in the ordering with randomly chosen players.
Figure S\ref{fig_noise} shows a setting like that of Figure 3
%\ref{fig_2}
of the main text, after we have corrupted the payoffs 
with a (quenched) noise set by $\sigma^2=1$, and reshuffled a proportion $f=25\%$ of players. Although the targeted punishment 
strategies lose some of their effectivity, there are still large regions of parameter space where their adoption 
leads to most players cooperating.

\begin{figure}[ht!]
\begin{center}
\includegraphics[scale=0.55]{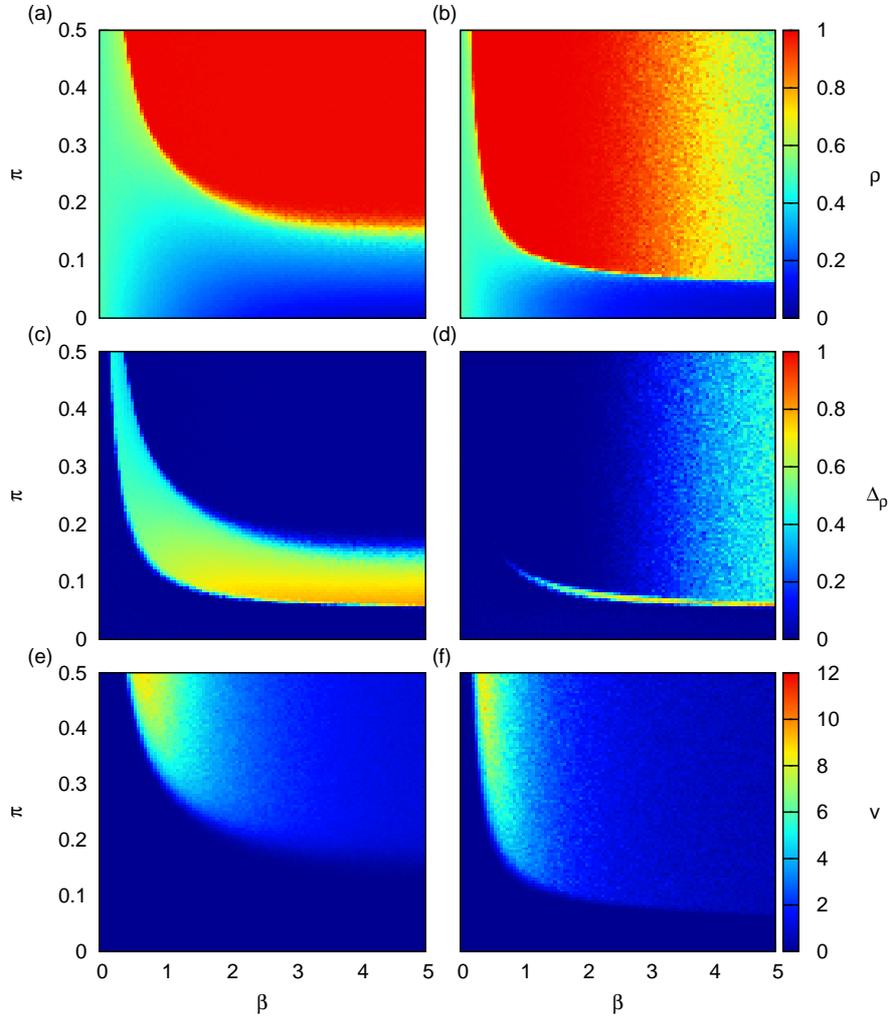}%{fig_noise_ap1_SI.eps}%{fig_2_SI.eps}%{fig_xscat.eps}%{fig_1_sketch.eps}
\end{center}
\caption{
As Figure S\ref{fig_ap1_tp} (all players punish), 
after the payoffs have been corrupted with a noise drawn from a Gaussian of mean zero and variance $\sigma^2=1$;
and a random proportion $f=25\%$ of players have had their positions in the ordering switched with other random players.
Punishment strategies are `single file' in the panels on the left [(a),(c) and (e)], and `groups' in those on the right [(b),(d) and (f)].
(a) and (b) Stationary proportion of cooperators, $\rho$.
(c) and (d) Difference between maximum $\rho$ maintainable [as displayed in Figure S\ref{fig_ap1}a], and the results of
Figures S\ref{fig_noise_ap1}a and S\ref{fig_noise_ap1}b, repectively.
(e) and (f) Speed $v=N/\tau$, where $\tau$ is the number of time steps required to achieve global cooperation, 
for the situations in Figures S\ref{fig_noise_ap1}a and S\ref{fig_noise_ap1}b, respectively.
}
\label{fig_noise_ap1}
\end{figure}

In Figure S\ref{fig_noise_ap1}, we carry out the same corruption of payoffs and of the ordering as in Figure S\ref{fig_noise}, but here 
we consider the situation where all players are punishers: $n_p=N$. Again, the strategies can be seen to be fairly robust under 
these conditions.

%IPCC SAR SYR (1996). Climate Change 1995: A report of the Intergovernmental Panel on Climate Change. Second Assessment Report of the 
%Intergovernmental Panel on Climate Change. IPCC.